\documentstyle[aps,preprint,eqsecnum]{revtex}

\begin{document}
\draft
\title{Effective Mass of Composite Fermions and Fermionic Chern-Simons Theory 
in Temporal Gauge}
\author{Yue Yu, Zhao-Bin Su and Xi Dai}
\address{Institute of Theoretical Physics, Academia Sinica, Beijing 100080,
China}

\maketitle
\begin{abstract}
The definitions of the effective mass of the composite fermion are discussed
for the half-filled Landau level problem. In a recent work, 
Shankar and Murthy show a finite effective mass of the composite fermion
by a canonical transformation while the perturbative calculation gives 
the logarithmic divergence of the effective mass at the Fermi surface.
We will emphasize that the different definition of the effective mass  
is related to the different physical processes. The finite one could be defined for 
any momentum of the composite fermion while  the divergence only appears
at the Fermi surface. We work with the standard Halperin-Lee-Read model but
in the temporal gauge. The advantage of this gauge to be employed
is that the finite effective mass could be determined in the Hartree-Fock
approximation. Furthermore, it is precisely equal to the result that 
Shankar and Murthy obtained which is well-fit with the numerical 
calculation from the ground state energy analysis and a 
semi-classical estimation. However, if we consider the random phase
approximation (RPA), one sees that the divergence of the effective mass of
the quasiparticle at the Fermi surface emerges again no matter that
we work on the temporal or Coulomb gauge. We develop an effective theory 
where the finite effective mass serves as a `bare' effective mass and show that
the same divergence of the RPA renormalized effective mass.
On the other hand, the correct behavior of the response functions 
in the small band mass limit could be seen clearly in the temporal gauge 
since there is a self-interaction among the magnetoplasmons.
\end{abstract}

\pacs{}


\section{Introduction} 

The fractional quantum Hall (FQH) metallic state at the half-filled Landau 
level was observed in 1989 \cite{Jiang}, which was a very important stage
attracting the theoretical research interesting from the odd denominator 
filling fractions to the even denominators \cite{HLR,KZ} since the FQH
effect (FQHE) was discovered \cite{Tsui}. There are two strategies to do 
theoretical research of the FQH metallic states: One is projecting 
the quantum states to the lowest 
Landau level (LLL) first and understanding the essential physics through 
the trial wave functions \cite{RR}. However, the quantum state space 
in the LLL is
tough to be dealt with in an analytical way. Most researches are restricted 
in the numerical simulation. Another approach is so-called the fermionic 
Chern-Simons theory \cite{FL}, which gives a field theoretical way to calculate the 
observed properties \cite{HLR,KZ}. Recently, Read has explained some relations 
between those two different ways \cite{Read}. In this paper, we would like to
express this kind of relations more explicitly.

In understanding the prominent plateaus of FQHE, an important concept is that
the FQHE of the original electrons in the external magnetic field yields
the integer quantum Hall effect (IQHE) of the composite fermions (CFs) 
which combines
the electron with an attached flux tube carrying even flux quanta 
\cite{Jain}. This concept is also applied to the FQH metallic states. However,
if one turns on a flux tube adiabatically, there will
always be a depletion of charge in the immediate vicinity of the location
of the flux tube. This is regarded as a vortex. The charge depletion was 
ignored in the fermionic Chern-Simons theory, which motivates Read \cite{Read} to improve 
the CF picture with an analogue to Laughlin's original quasiparticle notion
in which the zeros of the wave function reflect the charge depletion. 
According to the improved CF picture, a numerical calculation shows a 
very good overlap between Rezagy-Read trial wave function and the `exact' 
one for the small system \cite{RR}. Read also pointed
out that although the original electron's kinetic energy is completely 
quenched in the LLL, the Coulomb interaction that the vortex-electron pair
(i.e., CF) experiences induces an effective kinetic energy then an 
effective mass of the CF. Following this idea, we will give a semi-classical 
estimation of the effective mass which is very close to the numerical 
result for the small system which is based on an analysis of the  
ground state energies \cite{MdA}. 

The field theoretical description of the FQH metallic states is 
well-established by Halperin, Lee and Read (HLR) in their seminal paper 
\cite{HLR} (also see \cite{KZ}).
An important prediction in their work is that there is a CF Fermi surface 
which has been confirmed by a number of experimental works \cite{Exp}.
This implies that it is possible that some kind of the modified Fermi
liquid theory can be held at the low temperature. The central focus after
HLR's theory is whether the effective mass is divergent 
\cite{HLR,KFWL,AIM,KHM,GW,NW,Pol,SH,KWLS,KWL}. 
The essential results in the previous development are that the divergence 
of the effective mass is shown in the single-CF correlation function while the 
divergence can be cancelled in two-CF correlation functions. The latter 
is gauge invariant. Recently, there are two works 
\cite{CHY,SM} which have a finite answer in contrast to the divergent 
result. The work devoted by Chari, Haldane and Yang \cite{CHY} 
defines the effective mass in a gauge invariant way. However, 
although the effective mass defined in such a way is finite, there is no 
comparison
 to the numerical result and others because the effective mass 
that they calculation is cut-off dependent. The authors of ref.\cite{SM} 
provide a canonical transformation to the Hamiltonian of the 
fermionic Chern-Simons theory and find there are two mass scales related to  
the effective mass and the cyclotron motion frequency, respectively. 
The effective mass they obtained is finite and well-fit 
with the numerical result \cite{MdA}. It is no wonder that the effective  
mass defined in ref.\cite{CHY} is finite because it is gauge invariant.
However, notice that the effective mass obtained by Shankar and Murthy
somewhat is in the single particle sense. Then, one can ask that in what
physical processes we could observe the finite property of the effective
mass and otherwise the divergent property is shown.

We work in the standard HLR model with a different gauge choice. 
We work on the temporal gauge, whose advantage is that the mean-field 
state that the perturbative theory is based on is closer to the 
physical ground state. We deal with the theory in two steps. First,   
we would like to compare our result to the numerical result which 
is based on an analysis to the ground state energy \cite{MdA}.
So, we consider the approximation that a CF motion in an $N$ non-interacting
CF background. This is nothing but the Hartree-Fock approximation (HFA). 
After calculating the self-energy of the CF in a simple one loop digram, 
one sees that the renormalized effective mass is independent of the band
mass of the electron and precisely equal to the result obtained in the 
canonical transformation calculation \cite{SM}. Now, we could understand why
the effective mass calculated by the canonical transformation is
well coincided with the semi-classical estimation or the numerical
calculation based on the analysis of the ground state energy because
the HFA basically reflects the ground state property. The physical 
properties related to the ground state could be used to check  
the finiteness of the effective mass. Furthermore, we could understand why 
the energy gap gives the FQHE determined by the Coulomb interaction 
only because the effective Landau level gap is $\omega^*_c=\frac{e\Delta B}
{m^*c}$. We point out that this effective mass in the HFA is well-defined 
not only at the Fermi surface but also for any value of the momentum 
the CF carries. So, we can have an effective theory in which the  
mechanic kinetic energy is defined by the effective mass  while
the interaction between the CF and gauge field keeps no change. Namely, 
$m_b$ is not replaced by $m^*$ 
in the interaction vertex while the finite effective mass
serves as the `bare' mass in the kinetic term. 
Based on this effective theory, one can study the perturbative theory which is 
equivalent to the perturbation
directly starting from the original CF Hamiltonian. In the random phase 
approximation (RPA), the renormalized effective mass of quasiparticle shows a logarithmic
divergent behavior in consistence with HLR's result. This complex of the 
effective mass  is very similar to the vortex effective mass in HeII 
superfluid \cite{AN}. Moreover, it is easy to see that the high frequency  
gauge fluctuations in the RPA calculation, in fact, do not renormalize the 
HFA result, i. e., the RPA correction to the self-energy comes from
the low-frequency part of the gauge fluctuations.

Except the difficult status of the CF effective mass, another kind of
problem of the fermionic Chern-Simons theory is explored by Stern, Simon
and Halperin recently \cite{SSH}. In HLR's theory, one replaces the electron
band mass by a phenomenological effective mass, which causes, on the one hand
, Kohn's theorem is violated and the incorrect energy scale of the
response functions in the small band mass limit (equivalently, the high
magnetic field limit), on the other hand. The former problem is solved by
introducing a Fermi liquid parameter in HLR's theory while the latter 
one is claimed beyond the theory for which Stern, Simon and Halperin 
\cite{SSH} has to introduce an orbital magnetization attaching to the CF by 
hand. In fact, this magnetization could be induced from the interaction 
among the magnetoplasmons \cite{SM}, which is beyond to the RPA. 
In the meanwhile, the interaction among the
magnetoplasmons can not be renormalized, which relates to the magnetoplasmon
dispersion obeys Kohn's theorem. Those results could be shown perturbatively
in the temporal gauge which will be another contribution of the present
paper.

This paper is organized as follows. In Section II, we emphasize the physical
picture of the CF and estimate its effective mass in a semi-classical way.
In Section III, we review the fermionic Chern-Simons theory and work in 
the temporal gauge. The mean-field theory and the perturbative theory 
are formulated. 
In Section IV, the HFA and RPA calculations of the self-energy of  CF
are provided. In Section V, we define the effective masses of the CF according 
to the HFA and RPA self-energies respectively. One shows that the finiteness 
or divergence of the effective mass in the different situation.   
In Section VI, we discuss the response functions and their small band 
mass limit. The last section devotes our conclusions.

\section{ Physical picture of composite fermions}

The notion of the CF was first introduced by Jain in order to give the
prominent FQH states observed in experiment \cite{Jain}. However, 
in Jain's picture, the density variation in the immediate vicinity 
of the flux tube attached to an electron was not considered.
The fermionic Chern-Simons theory is based on this picture \cite{HLR}. 
Recently, Read \cite{Read} gave an improved
CF picture which considers the local environment change when 
a flux tube at the position of the extra electron is adiabatically turned on. 
According to Laughlin's quasiparticle notion, a quasihole can be obtained 
by acting the creation operator 
\begin{equation}
U(z)=\prod_{i=1}(z-z_i),
\label{HC}
\end{equation}
to the ground state wave functions. Here $z=x+iy$ is the complex variable
in a two dimensional plane. In the immediate vicinity of $z$, there is
a charge depletion because the zeroes of the wave function. Thus, the 
neutralizing uniform
positive background is naked in the vicinity of the local position while an 
equal magnitude but opposite sign charge accumulates at the boundary. 
We call this local environment change as a vortex at $z$. As well known, the 
vortex carrys a fractional charge $\nu e$ and has the fractional statistics
$\theta=\pi\nu$. Similarly, the operator $U(z)^{\tilde\phi}$ creates 
a $\tilde\phi$-fold vortex at $z$. Now adding an extra
electron near $z$, the $\tilde\phi$-fold vortex and the electron attract 
each other. Since all electrons lie on the LLL and then their
kinetic energies are quenched, the electron and the vortex
could form a bound state. This bound state, in Laughlin's case, is a boson 
with zero charge if $\nu=1/\tilde\phi$ 
because $\tilde\phi$ is odd then the vortex is a fermion. 
Read has noticed that for the case of even $\tilde\phi$, this vortex picture
has still a good chance to work. The trial wave function 
including a Jastrow-like factor \cite{RR}, say for $\nu=1/\tilde\phi=1/2$, is 
\begin{equation}
\Psi={\cal P}_{LLL}\det M \prod_{i<j}(z_i-z_j)^{\tilde\phi}e^{-\frac{1}{4
l_{1/\tilde\phi}^2}\sum_i |z_i|^2}.
\label{GSW}
\end{equation}
where $l_{1/\tilde\phi}=\sqrt{\hbar c/e B} $ is the magnetic length.
The matrix $M$ has elements that are essentially plane waves, $M_{ij}\sim
e^{{\vec k}_i\cdot{\vec r}_j}$. ${\cal P}_{LLL}$ projects all electrons to the
LLL. In this case, the bound state is a fermion with its net charge zero.
At low energies, we can consider such fermions as quasiparticles of the
system which can not condense to zero wave vector due to Pauli's principle.
So, a quasiparticle generally has its wave vector $\vec{k}$. In the LLL, this 
wave vector gives the separation of the electron from the vortex center.
To see this point, as Read noticed \cite {Read}, we consider the operator 
$\psi_e^\dagger(z) U(z)^{\tilde\phi}$ which creates a fermion. 
The quasiparticle with the wave vector $\vec{k}$ is created
by acting on the ground state with 
\begin{equation}
\int d^2z e^{i\vec{k}\cdot\vec{r}}
\psi_e^\dagger(z)U^{\tilde\phi}(z)\exp\{-\frac{1}{4l_{\tilde\phi}^2}|z|^2\}.
\label{CF}
\end{equation}
In the LLL, the 
operator $\bar z$ acts on a state in 
the Hilbert space likes $2\partial/\partial z$ \cite{GJ}. 
So the exponential $e^{ik \bar{z}}$ is like a translation operator, i.e.,
if $f(z_1, ...,z_j, ..., z_N)$ is the prefactor of a wavefunction, 
then we have 
\begin{equation}
e^{ik\bar z_j} f(z_1, ...,z_j, ..., z_N)
\sim f(z_1, ...,z_j+ik, ..., z_N).
\end{equation}
 The $j$-th particle is displaced by $ik=i(k_x+ik_y)$.
A quasiparticle with $k=0$ would have the electron exactly on the center of 
the vortex. So a quasiparticle with $k\not=0$ has the electron displaced 
by $|k|$ from the vortex center. This separation means that the electron and 
the vortex experience a potential $V(|k|)$ which is caused by the static
electric field of the other electrons. With the external magnetic field, 
the neutral fermion drifts along an equipotential line of $V(|k|)$ 
with a velocity $\sim \partial V(|k|)/\partial |k|$. 
Near the bottom of the potential, it will be quadratic, and the quasiparticle
has its effective mass $\sim (\partial^2 V(|k|)/\partial|k|^2)^{-1}$ in the
kinetic energy sense.

After the previous physical intuitive description, we could
estimate the effective mass of the CF semiclassically.  Assume a vortex
carrying two flux quanta centers in the origin and attaches to an electron 
located at $\vec{x}$. (We restrict our discussion to $\nu=1/2$. )
The Coulomb potential that the electron feels reads
\begin{equation}
V(\vec{x})=
\sum_{i=1}^{N} \frac{e^2}{\varepsilon |\vec{r}_i-\vec{x}|}=\int_D
d^2\vec{r}\rho(\vec{r})\frac{e^2}{\varepsilon |\vec{r}-\vec{x}|},
\label{poten}
\end{equation}
where $N$ is the electron number of the system and $\rho(\vec{r})$
is the electron density which tends to $\rho_0$, the average density
of the system, for sufficient large $r>>d_v/2$. Here $d_v$ stands for the
 size of the vortex. Hence, there is an attraction between the electron
and the vortex with the binding energy
\begin{eqnarray}
&&\int_Dd^2\vec{r}\rho(\vec{r})\frac{e^2}{\varepsilon |\vec{r}-\vec{x}|}
-\int_Dd^2\vec{r}\rho_0\frac{e^2}{\varepsilon |\vec{r}-\vec{x}|}
\nonumber\\&&= -\int_{D_0}d^2\vec{r}(\rho_0-\rho(\vec{r}))
\frac{e^2}{\varepsilon |\vec{r}-\vec{x}|},
\label{be}
\end{eqnarray}
where $D_0$ denotes the space range of the vortex. 
Because the electron kinetic energy is completely quenched in the LLL,
this energy binds the vortex and the electron that form a bound state
which we call the CF. In the many-CF system, we deal with the second
term on the left-hand side of the equality in (\ref{be}) as the interaction
between the CF and the neutralizing background and the first term as
the potential the CF experienced. By using the Legendre polynomials,
one can expand the potential as
\begin{eqnarray}
V(x)&=& \frac{e^2}{\varepsilon}\int_0^xrdr d\theta\rho(r)\sum_{l=0}
\frac{r^l}{x^{l+1}}P_l(\cos\theta) \nonumber \\
&+& \frac{e^2}{\varepsilon}\int_x^Rrdr d\theta\rho(r)\sum_{l=0} 
\frac{x^l}{r^{l+1}}P_l(\cos\theta),
\label{appV}
\end{eqnarray}
where $R$ is the radius of the system which is a macroscopic quantity.
Using the approximation employed by Laughlin in discussing the charge 
of the quasiparticle \cite{QHE}, one assumes there is no electron  
charge density in the regime of the vortex located (i. e., $\rho(r)=0$
for $r<d_v/2$) and  $\rho(r)=\rho_0$ out of this regime. Then the unit 
charge of the vortex means that $d_v/2=2l_{1/2}$. The effective mass is 
estimated by 
\begin{equation}
\frac{1}{m^*}=\frac{\partial^2 V(x)}{\partial x^2}\biggl|_{x=l_{1/2}}. 
\end{equation}
Plugging  (\ref{appV}) into the definition of the effective mass,
one has
\begin{equation}
\frac{1}{m^*}\approx 0.169\frac{e^2 l_{1/2}}{\varepsilon},
\end{equation}
which is very close to the value of $m^*=\frac{1}{6}
\frac{e^2 l_{1/2}}{\varepsilon}$ obtained through the canonical 
transformation \cite{SM}. Also, it is well-fitted with the numerical result 
$m^*=0.2\pm 0.02\frac{e^2 l_{1/2}}{\varepsilon}$ \cite{MdA}.  
The physical implication why those results of the effective mass calculation  
are so close is all those approaches to evaluate the effective mass are
essentially based on the analysis to the ground state properties.

To end this section, we argue why the fermionic Chern-Simons theory
is possible to describe the neutral object. First, Let us see the behavior of the quasiparticle when an applied electric field
is turned on. If we apply an external electric field to the system, 
the charges accumulated on the boundary with a lower electric potential move 
into the sample and the charges on the higher potential boundary move 
out the sample. This causes an electric current through the sample. 
In the bulk, unlike an ordinary neutral object which can not feel the 
electric field, the neutral fermions do response to the electric field.
For a neutral fermion with a wave vector $\vec{k}$, the separation of its 
electron from the vortex center varies to $|\vec{k}+\vec{q}|$. This implies
the potential varies to $V(|k+q|)$ and so  does the drift velocity. 
Therefore, the fermions have a neutral particle current in the electric field 
direction. Because of the one-to-one correspondence between the neutral 
fermion and one electron charge accumulated on the boundary, the neutral 
current strength of the fermions 
is equal to the electric current strength through the sample.  
Therefore, one can think the CF has one electron charge while
there are no charge accumulation on the boundary and no net magnetic field
acting on the CFs. In fact, the latter has been reflected in the
factor $\exp\{-\frac{1}{4l_{\tilde\phi}^2}|z|^2\}$ multiplied to the composite
fermion operator (\ref{CF}). 

If the external magnetic field slightly changes, the neutral quasiparticles 
also response to the change in the following sense.
If the filling factor is slightly away 
$1/\tilde\phi$, say $\displaystyle \frac{p}{\tilde\phi p+1}$ for a large $p$,
the CFs carry a net charge $-e^*=-\displaystyle\frac{e}{
\tilde\phi p+1}$. In the magnetic field $B$, the CF feels 
Lorentz' force 
\begin{equation}
\vec{F}=-e^*\vec{v}\times \vec{B}=-e\vec{v}\times \Delta\vec{B},
\label{LF}
\end{equation}
with $\Delta B=B/(\tilde\phi p+1)$. Again, the CF could be thought
carrying one electron charge. The CFs with charge $-e$ feel the
residual magnetic field $\Delta B$, which is what we will adopt. The ground state is 
a Fermi sea filled by the vortex-like composite fermions if $\Delta B\to 0$.

\section{Composite fermion in temporal gauge}

In this section, we take a field theoretical way to understand the CF and  
physics at $\nu=1/2$. Because it is very tough to direct construct a field
theory at the LLL, we will use the full electron field first. 
The LLL projection will be reflected in a truncation of momentum 
space.

\subsection{Hamiltonian and Lagrangian of Composite Fermions}

We start with a two dimensional interacting electron system which is 
placed in a uniform magnetic field $B$ perpendicular to the two dimensional 
plane in which there is a uniform positive background. One assumes that 
all electrons are polarized so that the spin degrees of freedom can be
ignored. For the two-body interaction potential $V$, the $N$-electron
Hamiltonian reads,
\begin{equation}
H_e=\frac{1}{2m_b}\sum_i\biggl[-i\hbar\nabla_i-\frac{e}{c}\vec{A}_i(\vec{x}_i)
\biggr]^2+\sum_{i<j} V(\vec{x}_i-\vec{x}_j),
\end{equation}
where the vector potential $\vec{A}$ is corresponding to the magnetic field
$B$ and $m_b$ is the band mass of the electrons. Hereafter, we will use the 
unit $\frac{e}{c}=\hbar=1$. Here we do not confine the electrons in the 
LLL. The attraction between the electrons and the uniform background is not
explicitly shown up. 

Following a common way, we make an anyon transformation
 \cite{LMW}. Writing the electron wavefunction $\Phi(z_1,...,
z_N)$ with $z_j=x_j+iy_j$, the position of the $j$-th electron, the 
transformation to the wavefunction reads
\begin{equation}
\Psi_{cs}(z_1,...,z_N)=\prod_{i<j}\biggl[\frac{z_i-z_j}{|z_i-z_j|}
\biggl]^{\tilde\phi}\Phi(z_1,...,z_N),
\end{equation}
where $\tilde\phi$ is an even number and then $\Psi_{cs}$ is the  
wavefunction of the Chern-Simons fermion. Here we distinguish the 
terminology {\it Chern-Simons fermion} to the CF for the reason explained
later. The corresponding Hamiltonian is given by  
\begin{equation}
H_{cs}=\frac{1}{2m_b}\sum_i\biggl[-i\nabla_i+\vec{A}_i
(\vec{x}_i)-\vec{a}_i(\vec{x}_i)\biggr]^2+\sum_{i<j} V(\vec{x}_i-\vec{x}_j),
\label{csh}
\end{equation}
where $\vec{a}$ is a statistical gauge potential
\begin{equation}
\vec{a}(\vec{x}_i)=\frac{\tilde\phi}{2\pi}\sum_{j\not=i}\frac{
\hat{z}\times(\vec{x}_i-\vec{x}_j)}{|\vec{x}_i-\vec{x}_j|^2}.
\label{sg}
\end{equation}
The statistical gauge field $\vec{a}$ induces a statistic magnetic field 
which reflects a constraint
\begin{equation}
b(\vec{x})=\nabla \times \vec{a}(\vec{x})=2\pi\tilde\phi\rho(\vec{x}).
\label{cst}
\end{equation}

The advantage of the CF picture is that the FQHE of the original electron
system yields the IQHE of the CFs \cite{Jain}. In the fermionic Chern-Simons
theory, it could be reached in the mean field theory,
\begin{equation}
\vec{a}_{MF}=\vec{A},
\end{equation}  
where we work in the symmetric gauge, i. e., 
$\vec{A}=(B/2)\hat{z}\times\vec{x}$. Around the mean field state, there is 
an important gauge fluctuation $\vec{A}-\vec{a}$.
Hereafter, we denote $\vec{a}$ as the fluctuation of the statistic field 
around the mean field. In this case, if we introduce the Chern-Simons 
fermion field $\psi_{cs}$ in the second quantization theory, 
then the corresponding Lagrangian reads,
\begin{equation}
L=\int d^2x \biggl[\psi_{cs}^\dagger(\vec{x},t)(i\partial_t-e a_0)\psi
_{cs}(\vec{x},t)+\frac{e}{2\pi\tilde\phi}a_0(\vec{x},t)\epsilon_{ij}\partial_i
a_j(\vec{x},t)\biggr]-H.
\end{equation}
Here the Hamiltonian has been taken in the form
\begin{equation} 
H=\int d^2x \frac{1}{2m_b}\biggl|(-i\nabla
+\vec{a}(\vec{x}))\psi_{cs}\biggr|^2+\frac{1}{2}
\int d^2xd^2x'\delta \rho(\vec{x})V(\vec{x}-\vec{x}')\delta\rho(\vec{x}').
\end{equation}
where $\delta\rho=\rho-\rho_0$.
Note that here we wrote down the Lagrangian in the Coulomb gauge 
$\nabla\cdot \vec{a}=0$. It is well-known that the theory has a gauge 
invariance corresponding to the gauge transformation of $\vec{a}$. Hence, 
the Lagrangian can also be written
in a gauge invariant form if we consider the bulk states only \cite{note1}.
Then, one can choose other gauge to deal with the system. For our case, 
an appreciate choice is so-called temporal gauge, i.e.,
\begin{equation}
a_0=0.
\end{equation}
Recall the transverse component of the Chern-Simons gauge field 
canonically conjugates to the longitudinal component, one has
\begin{equation}
[a_i,a_j]=\epsilon_{ij}.
\end{equation}
The Hamiltonian in temporal gauge reads
\begin{eqnarray}
H_{cf}&=&\int d^2x \biggl|(-i\nabla+\vec{a}(\vec{x}))
\psi_{cf}(\vec{x})|^2\nonumber\\
&+&\frac{1}{2}\int d^2x\int d^2x'\delta \rho(\vec{x})
\frac{e^2}{\varepsilon|\vec{x}-\vec{x}'|}\delta\rho(\vec{x}'),
\label{cfh}
\end{eqnarray}
where we have specified the interaction to be the Coulomb interaction.
The suffix `cf' implies that we will consider
the fermion field $\psi_{cf}$ in temporal gauge as the CF field. The gauge
transformation between $\psi_{cs}$ and $\psi_{cf}$ has been explicitly
shown in ref.\cite{SM}, which is just a normal one from the Coulomb gauge
to the temporal gauge (see below). The Lagrangian in the temporal   
gauge reads
\begin{equation}
L_{cf}=\int d^2x\biggl[ \psi^\dagger_{cf}(\vec{x},t)i\partial_t\psi_{cf}
(\vec{x},t)+\frac{e}{2\pi\tilde\phi}a_i(\vec{x},t)\epsilon_{ij}\partial_t 
a_j(\vec{x},t)\biggr]-H_{cf}.
\end{equation}

\subsection{Mean-Field State and Perturbative Theory}

We are going to deal with the variational ground state based on this 
mean-field consideration and the perturbative theory around the ground state.
One can rewrite the Hamiltonian (\ref{cfh}) as
\begin{equation}
H_{cf}=H_{0f}+H_{0a}+H_{i}+H_{ia},
\end{equation}
where
\begin{eqnarray}
H_{0f}&=&\frac{1}{2m_b}\int d^2 x |\nabla\psi|^2, \nonumber \\
H_{0a}&=&\frac{\rho_0}{2m_b}\int d^2 x(a_x^2+a_y^2)+\frac{1}{8\pi^2
\tilde\phi^2}\int d^2xd^2x'
\nabla\times\vec{a}(\vec{x})V(\vec{x}-\vec{x}')\nabla'\times\vec{a}
(\vec{x}'), \nonumber\\
H_{i}~&=&\int d^2 x \vec{a}\cdot \vec{j},\nonumber\\
H_{ia}&=&\frac{1}{4\pi\tilde\phi m_b}\int d^2x (\nabla\times \vec{a})
\vec{a}^2. 
\label{hdc}
\end{eqnarray}
Here $H_{0f}$ and $H_{0a}$ stand for the free Hamiltonian of the CF and 
the gauge fluctuation respectively. $H_i$ is the interaction between
the CF and the gauge field while $H_{ia}$ is the self-interaction of the
gauge field. The decomposition (\ref{hdc}) of the Hamiltonian is first  
given by Shankar and Murthy \cite{SM} who point out that  $H_{0a}$,
in fact, describes the magnetoplasmons of the theory and hence, 
the mean-field state wave function is given by the unprojection version of 
(\ref{GSW}). The recovering of the modular part of $(z_i-z_j)$ in this wave 
function is closely related to the gauge transformation $\psi_{cs}$
to $\psi_{cf}$. As shown by Shankar and Murthy \cite{SM}, the CF field
$\psi^\dagger_{cf}$, indeed, creates a Chern-Simons fermion 
$\psi^\dagger_{cs}$ and an associated hole. This agrees with Read's vortex
CF picture as we discussed in Sec. II.
The mean-field state energy, then, reads, 
\begin{equation}
E_0=\frac{N}{2}\hbar \omega_c(1+O(m_b/m^*)),
\label{e0}
\end{equation}
where $m^*\sim (e^2\l_{1/2}/\varepsilon)^{-1}$ will be regarded 
as the effective mass  of the CF as we will see in the coming sections.
Indeed, by using a canonical transformation that eliminates $H_i$
to the lowest order, Shankar and Murthy found that the original kinetic  
energy is quenched if one chooses a truncation of the wave vector
of the magnetoplasmon, $q<k_F$, while $1/m^*=e^2l_{1/2}/6\varepsilon$ 
serves as the effective mass. However, as we will see in the next section,
the lowest order elimination yields a simple HFA
to the CF self-energy. If we count the RPA contribution to the self-energy,  
the infrared divergence will appear again.  

The perturbative theory could start with to read out Feynman's rules from 
the Lagrangian. The free CF propagator (Fig. 1 (a)) is
\begin{equation}
G_0(k,\omega)=\frac{\theta(k-k_F)}{\omega-\epsilon_k+i0^+}+
\frac{\theta(k_F-k)}{\omega-\epsilon_k-i0^+},
\end{equation}
and the gauge fluctuation propagates like (Fig. 1(b))
\begin{equation}
D_0(q,\omega)=U,
\end{equation}
with
\begin{equation}                    
U^{-1}=\left( \begin{array}{cc}
         \displaystyle -\frac{\rho_0}{m_b} & {\displaystyle
          \frac{-i\omega}{2\pi\tilde\phi}}\\
          {\displaystyle\frac{i\omega}{2\pi\tilde\phi}}  &  \displaystyle
-\frac{\rho_0}{m_b}(1+\frac{e^2 q}{2\omega_c\varepsilon})\\
         \end{array} \right).
\label{GP}
\end{equation}
Here we have taken the $2\times2$ matrix description of the gauge propagator
with $D_{0\parallel,\parallel}=U_{11}$ and  $D_{0\perp\perp}=U_{22}$ and so on.
The suffices $\parallel$ and $\perp$ are corresponding to
the wave vector direction $\hat{q}$. The interaction vertex is shown as (Fig. 1 (c))
\begin{equation}
g_a=\frac{1}{m_b}((\vec{k}+\frac{\vec q}{2})\cdot\hat{q},
(\vec{k}+\frac{\vec q}{2})\times\hat{q}),
\end{equation}
while the gauge field self-interaction (Fig.1 (d)) is described by
\begin{eqnarray}
f_{222}&=&f_{211}
=\frac{i\vec{q}\cdot \hat{q}}{4\pi\tilde\phi m_b}, \nonumber\\
f_{122}&=&f_{111}=\frac{-i\vec{q}\times\hat{q}}{4\pi\tilde\phi m_b},.
\end{eqnarray}

\section{Calculations of CF Self-Energy}

In this section, we will calculate the CF self-energy in the HFA and RPA. 

\subsection{HFA}

We are going to calculate the CF self-energy within the HFA.
The CF (retarded) self-energy in the simple HFA, i.e., 
the one-loop approximation (Fig. (2)) is given by
\begin{eqnarray}
\Sigma^{*(0)}(k,i\omega_n)&=&-\frac{1}{(2\pi)^2\beta}\sum_{i\nu_n}
\int d^2qg_a{\cal D}_{0ab}(q,i\nu_n)g_b{\cal G}_0(k+q,i\omega_n+i\nu_n)
\nonumber\\
&=&\frac{4\pi\omega_c}{(2\pi m_b)^2}\int d^2q \biggl(k_\perp^2+(k_\parallel
+\frac{q}{2})^2(1+\frac{e^2q}{2\omega_c\varepsilon})\biggr)\nonumber\\
&\times&\frac{1}{2\omega(q)}\biggl[\frac{N_q+n_F(\xi_{k+q})}{i\omega_n+
\omega(q)-\xi_{k+q}}+\frac{N_q+1+n_F(\xi_{k+q})}{i\omega_n
-\omega(q)-\xi_{k+q}}\biggr],
\end{eqnarray}
where $\xi_k=\frac{k^2}{2m_b}-\mu$. $N_q$ and $n_F$ are the Bose and Fermi
factors with $\omega^2(q)=\omega_c^2(1+\frac{e^2q}{2\omega_c \varepsilon})$.
In the zero temperature limit and after an analytical continuation, one has
\begin{equation}
\Sigma^{*(0)}(k,\omega)=(-\frac{k^2}{2m_b}+\frac{k^2}
{2m^*})(1+O(\omega/\omega_c)).
\label{HFA}
\end{equation}
One would like to point out that we have taken a 
truncation when we integrate over the wave vector of the magnetoplasmon, 
$q<k_F$, which is what Shankar and Murthy used \cite{SM}. This truncation 
choice is
consistent with the correct mean-field ground state energy (\ref{e0})
and reflects the LLL projection.

\subsection{RPA}

Now, let us consider the further approximation. An easy and direct way going 
beyond Hartree-Fock's is the RPA which is through replacing the bare gauge 
propagator in Fig. 2 by the RPA one (Fig. 3),
\begin{equation}
D_R^{-1}=D^{-1}_0-K^0.
\end{equation}
Here, $K^0$ is the free CF response function (Fig. 4) instead of the full
response function in the RPA spirit. In the long wave length limit,
$q<<k_F$, the non-interacting response matrix can be explicitly written down
\begin{eqnarray}
K^0_{11}(q,\omega)&=&-\frac{\rho_0}{m_b}\biggl[1+2a^2-2a^3F(a)
+\frac{q}{k_F}(-\frac{3}{2}a+3a^3+(3a^2-3a^4)F(a))
\nonumber\\ 
&+&\frac{q^2}{k_F^2} 
(\frac{1}{2}-\frac{8}{3}(a^2-a^4)+(-\frac{3}{2}+4a^3-\frac{8}{3}a^5)F(a))
\nonumber \\
&-&i(a^3\tilde{F}(a)+\frac{q}{k_F}(3a^2\tilde{G}(a))+\frac{q^2}{k_F^2}
(\frac{3}{2}a-4a^3+\frac{8}{3}a^5)\tilde{F}(a))\biggr],\nonumber\\
K^0_{22}(q,\omega)&=&-\frac{\rho_0}{m_b}\biggl[1-2a^2+2aG(a)
+\frac{q}{k_F}(\frac{5}{2}a-3a^3+(3a^2-1)G(a))\nonumber\\
&+&\frac{q^2}{k_F^2}(\frac{2}{3}(4a^2-1)(a^2-1)-\frac{2}{3}(4a^2-3)aG(a))
\nonumber\\
&-&i(2a\tilde{G}(a)+\frac{q}{k_F}(1-3a^2)\tilde{G}(a)-2a(1-\frac{4}{3}
a^2)\tilde{G}(a))\biggr],
\end{eqnarray}
where $a=\frac{\omega m_b}{k_F q}-\frac{q}{2k_F}$ and the functions
$$
G(a)=\biggl\{\begin{array}{ll}
             \sqrt{a^2-1},& {\rm for}~~~ a>1;\\
             0,&{\rm for}~~~ a<1,
             \end{array}
$$
$$
\tilde{G}(a)=\biggl\{\begin{array}{ll}
                     \sqrt{1-a^2},&{\rm for} a<1;\\
                     0,&{\rm for } a>1,
                     \end{array}
$$
and $F(a)=(G(a))^{-1}$, $\tilde{F}(a)=(\tilde{G}(a))^{-1}$.

In the limit $a>>1$, one has $K^0_{11}\sim K^0_{22}\sim O(1/a^2)$. This 
means that the bare gauge propagator is not renormalized in the limit
$\omega>> v_F q$. In the opposite limit, $\omega<<v_Fq$, one has
\begin{eqnarray}
{\rm Re}K^0_{11}&\simeq& \frac{\rho_0}{m_b}(-1+2a^2+\frac{5}{2}\frac{q}{k_F}a
+\frac{2}{3}\frac{q^2}{k_F^2}),\nonumber\\
{\rm Re}K^0_{22}&\simeq&\frac{\rho_0}{m_b}(-1-2a^2-\frac{3}{2}\frac{q}{k_F}
a+\frac{1}{2}\frac{q^2}{k_F}),\nonumber \\
{\rm Im}K^0_{11}&\simeq& \frac{\rho_0}{m_b}(-2a^3-\frac{3}{2}\frac{q}{k_F}
a^2+\frac{1}{2}\frac{q^2}{k_F^2}a),\nonumber\\
{\rm Im}K^0_{22}&\simeq &-\frac{2\rho_0\omega}{k_Fq}.
\end{eqnarray}
There are the terms of order $O(1)$ in the real part of the response 
functions, which renormalizes the bare gauge propagator dramatically. 
Namely, the RPA corrected gauge propagator in the low-frequency limit
is given by
\begin{equation}
D_R^{-1}=\left( \begin{array}{cc}
        \displaystyle   -\frac{\rho_0}{m_b}(-2a^2-\frac{3qa}{2k_F}+\frac{q^2}{2k_F}) & 
          {\displaystyle
          \frac{-i\omega}{2\pi\tilde\phi}}\\
          {\displaystyle\frac{i\omega}{2\pi\tilde\phi}}  & \displaystyle 
-\frac{\rho_0}{m_b}(2a^2+\frac{5qa}{2k_F}+\frac{2q^2}{3k_F^3}
+\frac{e^2 q}{2\omega_c\varepsilon})+i\frac{2\omega m_b}{k_Fq}
         \end{array} \right).
\end{equation}
Note that there is a pole of $\det D_R^{-1}$ at 
\begin{equation}
\omega\simeq -i\frac{q^2 e^2}{4k_F\varepsilon},
\end{equation}
which exactly recovers what has been seen in the Coulomb gauge \cite{HLR}.  

Using the RPA corrected gauge propagator instead of the bare one, 
we can  calculate the RPA corrected CF self-energy (Fig. 5), which reads
\begin{equation}
\Sigma^*_R(k,i\omega_n)=-\frac{1}{(2\pi)^2\beta}\sum_{i\nu_n}
\int d^2qg_a{\cal D}_{Rab}(q,i\nu_n)g_b{\cal G}_0(k+q,i\omega_n+i\nu_n).
\end{equation}
In the zero temperature limit, the sum over the frequency tends to 
the integration and
\begin{eqnarray} 
\Sigma^*_R(k,i\omega)&=&-i\int \frac{d^2q}{(2\pi)^2}\frac{d\nu}{2\pi i}
\frac{1}{m_b^2}\frac{1}{i\nu+i\omega-\xi_{k+q}}\nonumber\\
&\times&\frac{4\pi\omega_c(1+4\pi K_{22}^0/\omega_c+e^2q
/2\omega_c\varepsilon)(k_1+q/2)^2+4\pi\omega_c k_2^2(1+
4\pi K_{11}^0/\omega_c)}{\nu^2+\omega_c^2(1+4\pi K_{11}^0/
\omega_c)(1+4\pi K_{22}^0/\omega_c+e^2q/
(2\omega_c\varepsilon)}.
\end{eqnarray}
Comparing to the Hartree-Fock case, there is one more
pole,  $\nu\sim -\frac{q^2 e^2}{4k_F\varepsilon}$, in the integrand.
Finally, the CF self-energy in the RPA reads \cite{HLR}
\begin{equation}
\Sigma^*_R(k,\omega)\simeq
-A\epsilon\ln(B\epsilon)+iC\epsilon+\Sigma^{*(0)}(k),
\label{RPA}
\end{equation}
where $A$, $B$ and $C$ are positive constant and $\epsilon=\omega-\mu$.

\section{Effective mass and effective theory}

\subsection{`Bare' Effective Mass} 

Now we are going to discuss the effective mass of the CF. The RPA 
result to the CF self-energy shows that we can decomposite the gauge
fluctuation into the high and low frequency parts. If the low-frequency 
gauge fluctuations are somehow suppressed, the high-frequency parts 
do not affect the HFA result because the bare gauge propagator is 
not renormalized by the high-frequency gauge fluctuations. So, we deal  
with the problem in two steps by switching off the low-frequency
gauge fluctuations first and then turning it on. If the low-frequency
gauge fluctuations is switched off, the self-energy is given by the 
HFA calculation (\ref{HFA}). The effective dispersion of the CF is defined
by 
\begin{equation}
\frac{k^2}{2m^*}= \frac{k^2}{2m_b}+\Sigma^{*(0)}(k).
\end{equation}
Since the HFA self-energy is frequency-independent,
the effective mass can be read off
\begin{equation}
\frac{1}{m^*}=\frac{e^2l_{1/2}}{6\varepsilon}.
\end{equation}
This is just the result obtained in ref.\cite{SM} by the canonical 
transformation method. Notice that this effective mass of the CF  
is not only well-defined at the Fermi surface but also to any value   
of the momentum. Furthermore, the high-frequency gauge fluctuations
do not affect the gauge propagator. Therefore, we can write down an effective
theory with the effective Hamiltonian
\begin{equation}
H_{\rm eff}=H_{0f}^*+H_{0a}+H_i+H_{ia}, 
\label{eh}
\end{equation}
where $H_{0a}$, $H_i$ and $H_{ia}$ are defined as before ( 
see (\ref{hdc}) ) and $H_{0f}^*$
is the free Hamiltonian of the CFs with the effective mass $m^*$. Our first
step gone forward in a while ago yields switching off the interaction $H_i$ 
and $H_{ia}$. Now, let us go away from the exact half-filled case. 
Then a residual magnetic field $\Delta B$ is turned on. 
In terms of the minimal coupling to the residual vector potetial $\Delta 
\vec{A}$, the effective `free' CF Hamiltonian $H^*_{0f}$ is replaced by 
\begin{equation}
H^*_{Af}= \frac{1}{2m^*(\nu)}\int d^2 x |(-i\nabla+\Delta \vec{A})\psi|^2,
\end{equation}
where $m^*(\nu)$ is a $\nu$-dependent effective mass. The dependence of the
filling fraction in  $m^*(\nu)$ comes from: 1) as $\nu_0=1/2\to \nu$, the
cyclotron frequency $\omega_c(1/2)\to \omega_c(\nu)$. This leads to 
$m^{*-1}\to m^{*-1}(\nu)\sim C_{1/2}e^2l_{\nu}/\varepsilon$ with $C_{1/2}=1/6$. 2)
The factor $C_{1/2}$ also could be varied as $\nu$. For instance,
for the small $\Delta B$, a term in Hamiltonian proportional to
$\Delta\vec A\cdot \vec j$ could modify $C_{1/2}\to C_\nu$.
So, in general, $m^{*-1}(\nu)\sim C_\nu e^2l_\nu/\varepsilon$ with $C_\nu$ in the
range $0.2\sim 0.4$ \cite{HLR,MdA}. This result qualitatively
agrees with the various experiments \cite{Du,Leadley}. 
In this sense, this effective mass defined in the HFA could be thought as 
a `bare' effective mass. 

According to Jain \cite{Jain},
all $p$ Landau levels of the CFs are fully filled if   
\begin{equation}
\Delta B=\frac{B}{2p+1}.
\end{equation}
The electron filling fraction, then, is given by
\begin{equation}
\nu=\frac{p}{2p+1},
\end{equation}
which manifests that the IQHE of the CFs yields the FQHE of the electrons.
Furthermore, the Landau gap of FQHE is given by
\begin{equation}
E_{\nu}=\omega_c^* =\frac{B}{|2p+1|m^*}.
\end{equation}
One sees that the Landau gap of FQHE is totally determined by 
the interaction as one expects.

To summarize this subsection, we see that the effective mass $m^*$
can be thought as the `bare' mass of CF in the effective theory and it
essentially  is the result of the mass renormalization in the
HFA. Physically, the HFA describes the probe CF interacting
with the $N$-CF background without considering the vacuum fluctuation. This
implies that the HFA basically reflects the ground state behavior of the 
CF system. So, it is no wonder why the `bare' effective mass calculated
here is so close to the semi-classical estimation value of the effective 
mass (see Section II) and its numerical calculation value based on the
ground state energy analysis \cite{MdA}.

\subsection{Divergent Effective Mass and RPA}

In the last subsection, we have switched off the low-frequency gauge
fluctuations and see the `bare' effective mass is finite. In the case 
of the FQHE with the odd denorminator filling fraction, it is a good
approximation to neglect the low frequency gauge fluctuations 
because we have a finite Landau gap. However, as $\nu$ approaches to $1/2$, 
i. e., all Landau levels of the CFs are fully filled ($p\to\infty$), the  
Landau gap tends to vanish. The low-frequency gauge fluctuation may play 
an important rule in the theory. The perturbative theory associated with the
effective Hamiltonian (\ref{eh}) could begin with the following 
Feynman's rules. The effective CF propagator reads
\begin{equation}
G^*_0(k,\omega)=\frac{\theta(k-k_F)}{\omega-\epsilon^*_k+i0^+}+
\frac{\theta(k_F-k)}{\omega-\epsilon^*_k-i0^+},
\end{equation}
where $\epsilon^*_k=\frac{k^2}{2m^*}$. Others keep the same as those in 
Section III. Especially, one would like to emphasize that the band mass in 
the interaction vertices are not renormalized. A similar calculation 
of the CF self-energy in the RPA gives rise to
\begin{equation}
\Sigma^{\rm eff}_R(k,\omega)\simeq
-A\epsilon\ln(B\epsilon)+iC\epsilon+...,
\label{ese}
\end{equation}
which is essentially the same as the original RPA result except the omitted
terms are only weakly $k$-dependent. The effective mass $m^*_{RPA}$, 
in the present case, is defined by 
\begin{equation}
\frac{m^*}{m^*_{RPA}}=\frac{1+\partial \Sigma^{\rm eff}_R/\partial
\epsilon_k^*}{1- \Sigma^{\rm eff}_R/\partial \omega}\biggr|_
{\omega=\mu}.
\end{equation}
That is, the effective mass in the RPA is logarithmic divergent at the 
Fermi surface
\begin{equation}
m^*_{RPA}\sim m^*A|\ln(k-k_F)|.
\label{dvi}
\end{equation}
This is completely consistent with the effective mass obtained by using 
the perturbative theory developed in Section III. Eq.(\ref{RPA}) gives
the effective mass in the RPA through
\begin{equation}
\frac{m_b}{m^*_{RPA}}=\frac{1+\partial \Sigma^*_R/\partial
\epsilon_k}{1- \Sigma^*_R/\partial \omega}\biggr|_
{\omega=\mu},
\end{equation}
which also shows the divergence of the effective mass exactly like 
(\ref{dvi}). Comparing to the effective mass calculation in the Coulomb
gauge, one sees that the CF effective mass in the RPA has the identical
divergent form . 

Consequently, we have seen that whether the effective mass diverges is 
dependent on how strongly the low-frequency gauge fluctuations can affect the
propagation of the CF. The gauge fluctuation  here in fact reflects the 
tranverse current fluctuations of the system. This implies that if the CF 
propagation in a physical process does not response the long-time 
transverse current fluctuation, the finiteness of the effective mass is 
shown and otherwise it tends to divergence.

\section{Response functions in the small band mass limit}

In the previous discussions, we omit the contribution from the self-interaction
among the gauge fluctuations because its high energy behavior. However, in the
small band mass limit, this interaction $H_{ia}$ has a dominating 
contribution to the response functions since
the coupling constant $1/m_b$ is not renormalized. In a recent paper,
Stern, Simon and Halperin recognize another difficulty of the fermionic
Chern-Simons theory \cite{SSH}. Namely, the energy scale of the response functions to the 
external field is not proper in the small band mass limit or equivalently 
the high magnetic field limit. They claimed that to solve this problem has
beyond the fermionic Chern-Simons theory. Then, a magnetization has to be 
attached to the CF by hand. After this phenomenological attachment, 
the current is separated into a magnetization current associated with the
cyclotron motion of the electrons and a transport current associated with the
guiding center motion. The cyclotron motion of the electron is not 
renormalizable so that the magnetization current dominates in the 
small band mass limit, which gives the correct  energy scale of the 
response functions in the small band mass limit. Here, we will see that
this magnetization of CFs does not need to be attached  by hand and
is self-included in the theory. $H_{ia}$ supplies this magnetization
while the rest in the Hamiltonian (\ref{hdc}) describes the transport  
behavior of the theory.

The current in the present theory is given by
\begin{equation}
j_a=K_{ab} A_b, 
\end{equation}
where $K_{ab}$ are the full response functions. In the fermionic Chern-Simons 
theory, one replaces $K_{ab}$ by the RPA response functions 
\begin{equation}
K^R=K^0(1-UK^0)^{-1}.
\end{equation}
In fact, this approximation gives the transport current 
\begin{equation}
j^{tran}_a=K^R_{ab}A_b,
\end{equation}
since the gauge propagator is not renormalized in $\omega\gg v_Fq$. We see 
that the RPA result satisfys Konh's theorem. Namely, the pole 
of $K^R$ is at
\begin{equation}
\omega=\omega_c(1+\frac{e^2 q}{2\omega_c\epsilon}+O(\frac{q^2}{m_b}))^{1/2}.
\end{equation}
Notice that the effective
mass we obtained in the last section is related to the kinetic energy of the 
quasiparticle only so that in a further correction beyond the RPA, only 
$O(\frac{q^2}{m_b})$ is renormalized to $O(\frac{q^2}{m^*})$ but   
the leading term is not. This shows non-renormalizable of the
cyclotron frequency and the satisfaction of Konh's theorem. 

What we omitted in the proceeding treatment is the contribution from the 
self-interaction among the magnetoplasmons,
$H_{ia}$. In the lowest order, this interaction Hamiltonian devotes
a bubble of the gauge fluctuation like Fig. 6 to the response functions. 
This lowest order 
approximation yields rewriting $H_{ia}$ as \cite{SM}
\begin{equation}
H_{ia}=\frac{1}{2m_b}\int d^2 x \biggl[\frac{1}{4\pi}\nabla\times \vec{a}
\biggr] 4\pi \rho(\vec{x}).
\end{equation}
This contributes to the total current a cyclotron part 
\begin{eqnarray}
&&\vec{j}^{mag}=\mu_b\hat{z}\times \nabla\rho+{\rm fluctuations},\nonumber\\
&&{\rm or} \nonumber\\
&&j^{mag}_\perp(q,\omega)\simeq i\mu_bq\rho(q,\omega),~~~ j^{mag}_\parallel=0.
\end{eqnarray}
where $\mu_b=\frac{e\hbar}{2m_bc}$ is Bohr's magnet. 
This is just the magnetization current introduced in the Stern, Simon and 
Halperin's paper \cite{SSH}. 

Now, we are going to the response function in the small band mass limit.
First, we recall the current conservation,
\begin{equation}
j_\parallel(q, \omega)=\frac{\omega}{q}
\rho(q,\omega). 
\label{cont}
\end{equation}
In addition, the gauge transformation from the Coulomb 
gauge to the temporal gauge is given by 
\begin{equation}
A^T_\perp=A^C_\perp,~~ A^T_\parallel=\partial_\parallel \Lambda,~~ \Lambda=
-\int dt A_0.
\end{equation}
This gives 
\begin{equation}
A^T_\parallel(q,\omega) =\frac{q}{\omega} A_0 (q,\omega).
\label{CoTmp}
\end{equation}
Then, we can have the following relations between the
response functions:
\begin{eqnarray}
K_{00}(q,\omega)&=&\frac{q^2}{\omega^2}K_{11}(q,\omega), \nonumber\\
K_{02}(q,\omega)&=&\frac{q}{\omega}K_{12}(q,\omega).
\label{relation}
\end{eqnarray}

Using equations (\ref{cont}), (\ref{CoTmp}) and (\ref{relation}), one sees
that the magnetization current contributes to the response functions like
\begin{eqnarray}
\delta K_{21}&=&i\mu_bq \frac{q}{\omega} K_{11}, \\
{\rm or} \nonumber &&\nonumber\\
\delta K_{20}&=&i\mu_b q K_{00}.
\end{eqnarray}
and $\delta K_{11}=\delta K_{22}=0$. This gives the correct 
energy scale of the response functions in the small band mass limit \cite{SSH},
\begin{equation}
\lim_{m_b\to 0}\frac{K_{20}}{K_{00}}=i\mu_b q.
\end{equation}

In conclusion, we recover the correct small band mass limit of the response 
functions due to considering the self-interaction among the magnetoplasmons
which has been neglected in the previous literatures relating to the perturbative
theory of the fermionic Chern-Simons theory in the clue of HLR. It appears
because the magnetic current actually is devoted by Feynman's diagram 
(Fig. 6) beyond the RPA as we have seen.

\section{Conclusions} 

To summarize this work, we have seen that the physical picture of the CF 
based on the bound state of the electron and the attached vortex can be 
realized in a field theoretical way, the fermionic Chern-Simons theory, 
especially if the temporal gauge is taken into account. One found that there 
are two CF effective masses  which related to the different physical processes.
The finite effective $m^*$ is applied to the processes where the low-frequency 
gauge fluctuations become not important while the divergent one is 
associated with the low-frequency gauge field fluctuations dominate.
Furthermore, one found that the finite effective mass works for arbitrary
CF momentum but the divergent one could only well-defined near the Fermi 
surface. Therefore, we define an effective theory by using the finite 
effective mass as the `bare' mass while keeping the interaction terms
do not change. The RPA calculations either by directly perturbative theory or
by the effective theory showed the same logarithmic divergence of the
renormalized effective mass.

The other advantage of using the temporal gauge to the fermionic Chern-Simons
field theory is that one can somewhat go beyond the RPA of the perturbative 
theory. An example is the calculation of the total current. The RPA could
only give its transport part while the magnetization part comes from the 
self-interaction among the magnetoplasmons which has been ignored in 
the perturbative theory in the Coulomb gauge. So, we solved the problem about
the incorrect energy scale of the response functions in the small band mass 
limit.
   
So far we have partially cleared two puzzles in the fermionic Chern-Simons 
theory: the effective mass and the response functions in the small band mass 
limit. It would be interesting to relate our theory in the temporal gauge to 
various results in the Coulomb gauge. We will leave those relationships 
to the further work. Also the disorder problem is not involved in this paper.

The authors are gradeful Ping Ao to remind us the similar situation of
the effective mass in $HeII$ superfluid. They also thank V. J. Goldmann
to point out a citation error in the references.
This work wass supported in part by the NSF of China. One of us (Y. Y.)
was also supported by Grant LWTZ-1298 of Chinese Academy of Science.


\newpage

\centerline{\bf FIGURE CAPTIONS}

Fig. 1: Feynman's rules. The solid lines are the CF propagator and the dashed
lines are the gauge propagator.

Fig. 2: Hartree-Fock's self-energy of the CF.

Fig. 3: The RPA propagator of the gauge field.

Fig. 4: Free CF polarized diagram.

Fig. 5: The RPA self-energy of The CF.

Fig. 6: One-loop diagram of the self-interaction of the gauge field.

\end{document}